\documentclass[epj]{webofc}
\usepackage[utf8]{inputenc}
\usepackage[varg]{txfonts}   
\usepackage{booktabs}
\usepackage{xcolor}
\definecolor{darkred}{rgb}{0.4,0.0,0.0}
\definecolor{darkgreen}{rgb}{0.0,0.4,0.0}
\definecolor{darkblue}{rgb}{0.0,0.0,0.4}
\usepackage[bookmarks,linktocpage,colorlinks,
    linkcolor = darkred,
    urlcolor  = darkblue,
    citecolor = darkgreen]{hyperref}
\wocname{EPJ Web of Conferences}
\woctitle{Lattice2017}
\begin{document}
%
\selectlanguage{english}
\title{%
Wavelets and Lattice Field Theory.
}
\author{%
\firstname{Herbert} \lastname{Neuberger}
\inst{1}
\fnsep\thanks{Acknowledges partial support by the NSF under award PHY-1415525.  
Speaker, \email{neuberg@physics.rutgers.edu}.}}

\institute{%
Department of Physics and Astronomy, Rutgers University, Piscataway, NJ 08854, U.S.A
}
\abstract{%
When continuous fields are expanded in a wavelet basis, a D-dimensional continuum action becomes a
(D+1)-dimensional lattice action on the naively discretized Poincare-patch coordinates of an Euclidean AdS(D+1). New possible criteria for acceptable actions open up.   
}
\maketitle
\section{Introduction}\label{intro}

Using wavelets, a continuum action can be rewritten as an action of lattice fields. The path integral is over these new variables. At the classical level, gauge theories expressed in terms of Lie Algebra valued connections and also supersymmetric theories are well defined on this lattice. So long as the lattice remains infinite the lattice representation of the continuum action is exact. A continuum $D$ dimensional field theory is turned into a $D+1$ lattice field theory. 

The IR and UV problems look more symmetrical in this lattice form. UV and IR cutoffs are defined by limiting resolution. The index range on the $D+1$-axis is then bounded from above and from below respectively. The remaining directions remain infinite and $D$-space is not compactified.

\section{Resolution or Resolving-Power}\label{sec-1}

The extra dimension of the lattice separates resolutions. The intent is clear but the concept is not yet sharply defined. We start with a general square-integrable field $\Phi({\vec x})$. It has details on all scales. 

We wish to decompose $L_2$ ($L_2(R^D)$) into 
contributions of limited bandwidth slices in momentum space. The bands are deformations of spherical slices, equally spaced logarithmically. The slices fit together like onion shells. Each slice generates a subspace of the Hilbert space $L_2$, $W_j$, and the middle of the slice is at distance $2^j$ from the origin. All the $W_j$ spaces are infinite dimensional and mutually orthogonal in the inner $L_2$ product:
\begin{equation}
L_2=\oplus_{j\in Z} W_j
\end{equation}
Elements of $V_j=\oplus_{j' < j} W_{j'}$ consist of fields $\Phi\in L_2$ whose resolution is limited by $j$; they carry information down to finest length scales $2^{-j}$, but not finer. We require $V_\infty=L_2$ and $V_{-\infty}=\{0\}$. By construction, 
\begin{equation}
\{0\}\subset\cdots V_{-j} \subset V_{-j+1} \cdots V_{-1}\subset V_0\subset V_1\subset \cdots V_j\subset V_{j+1} \cdots
\subset L_2
\end{equation}
A specific construction of the $V_j$ defines a resolution. The $W_j$'s can be constructed from the
$V_j$ by solving for the orthogonal complement of $V_j$ in $V_{j+1}$:
\begin{equation}
V_{j+1}=V_j\oplus W_j
\end{equation}
The tower-like structure of the $V_j$ spaces is common to multi-grid and finite elements. Each
larger space $V_j$ contains finer detail than its predecessor. 

The term resolution reflects variability at some scale and is made precise by imposing at all $j$ the following scaling relation:
\begin{equation}
\Phi({\vec x})\in V_j \Leftrightarrow \Phi({2\vec x})\in V_{j+1}\label{scalinginclusion}
\end{equation}
$V_{j+1}$ is $2^D$ times ``larger'' than $V_j$. 

\section{Wavelets: what are they?}

We wish to find orthonormal bases for the $W_j$ (equivalently for the $V_j$) whose elements, in addition to being dominated by the contribution of one momentum shell in Fourier space, are also dominated by the contribution from some region in real space. The phase-space cell whose contribution dominates a given basis element has dimensions limited by the uncertainty principle which is obeyed optimally. 

The $j$-slices are scaled by powers of 2 and in each slice the basis elements are obtained by shifts of order $2^{-j}$. The relevant continuous symmetry is under the group of transformations ${\vec x}\to a{\vec x}+{\vec b}, a>0$. We first consider a set of coherent states made by applying all group elements to one special state $\psi({\vec x})$. This set will be over-complete and an orthonormal spanning subset is extracted,  labeled by discrete values of $a$ and ${\vec b}$. These values are identified as sites on a graph embedded in $D+1$ dimensional space. This is simplest in $D=1$. For $D>1$ one needs a finite set of special states $\psi^\alpha ({\vec x}),\; \alpha=1,\cdots,2^D-1$. $\alpha$ is viewed as an internal index. The $\psi^\alpha$'s are constructed from products of functions special to $D=1$. We substitute $\Phi({\vec x})$ in the local continuum action by its expansion in the $L_2$ basis elements $\psi^\alpha_{j{\vec n}}({\vec x})$ and integrate over ${\vec x}$. The coefficients in the expansion are the new path integration variables. 

\section{Locality \cite{resolvingpower}}\label{mainpoint}

What kind of locality properties in the expansion coefficients $\Phi^\alpha_{j{\vec n}}$ will the $(j,{\vec n})$-lattice action have? That one will have some form of locality in ${\vec n}$ in each shell is expected, but locality in $j$ is less obvious. In all familiar field theories the RG framework works, and its main ingredient is the impact of one momentum slice on the following one. Invariance under shifts in $j$ corresponds to a $j$-dependence reminiscent of the $AdS/CFT$ correspondence and deconstructed versions of the warp factor of Randall-Sundrum scenarios. One can think about the lattice as embedded in $D+1$ AdS space in typical coordinates, as evidenced by the $2^j$ factors. Only in the IR and/or the UV does one expect full restoration of scale invariance after quantization. 

The main point of \cite{resolvingpower} was to take the lattice action of the phase-space cell basis as a starting point and make it local in $j$ by decree. This would correspond to a continuum action which has some non-locality. We learned that some forms of non-locality do not cause expulsion from the basins of attraction of desirable fixed points from the study of lattice chiral fermions. Maybe one can get out of the fine-tuning bind associated with the Higgs mass by adopting a fundamental principle of some form of locality of the phase-space cell $(j,{\vec n})$-lattice action, instead of the space-time one.

\section{More details about wavelets}\label{moredetails}

There is a large literature on wavelets. I listed only books in \cite{waveletsbooks}. More references can be found in \cite{resolvingpower}. It is standard in the field to start the construction of wavelets in one dimension.

\subsection{Wavelets in $D=1$}

Wavelets have been looked at by lattice people and others physicists before me \cite{waveletsqft}. Wilson used some form of wavelets when he laid the foundations of the RG several years before wavelets were invented. 

The objective is to construct a useful basis in a space of square integrable functions, each associated with one phase-space cell.  (For $D>1$ there could be several basis elements per cell.)
The best known method starts by finding a special scaling function $\varphi(t)$ whose translates span $V_0$. $\varphi$ is concentrated in the phase-space cell $(j=0,n=0)$. All of $V_j$ is generated by $\varphi$. The associated $W_j$ are then defined by a wavelet $\psi$ built out of $\varphi$. 

We require that the set $\{\varphi_{0n}(t)\equiv\varphi(t-n)\}_{n\in Z}$ be orthonormal. 
This condition simplifies in Fourier space. I define Fourier transform and its inverse by:
\begin{equation}
{\hat\varphi(\xi)}=\int dt e^{-2\pi i \xi t}\varphi(t)~~~~~~~\varphi(t)=\int d\xi e^{2\pi i t\xi}\hat\varphi(\xi)
\end{equation}
Using the freedom to rescale the
argument of $\varphi$ and $\varphi$ itself while maintaining $\|\varphi\|^2=1$, we set $\int dt \varphi(t)=1=\hat\varphi(0)$. Orthonormality is equivalent to
\begin{equation}
\sigma_\varphi (\xi)\equiv\sum_{n\in Z} |{\hat\varphi(\xi-n)}|^2=1\label{orthonormality}
\end{equation}
Indeed, decomposing the integral over $\xi\in R$ in the inner product below into a sum of integrals over segments of unit length gives:
\begin{equation}
\langle \varphi_{00},\varphi_{0n}\rangle = \int_0^1 d\xi \left [ \sum_{k\in Z} |\hat\varphi(\xi-k)|^2 \right ] e^{2\pi i n\xi}
\end{equation}
If the left hand side of eq. (\ref{orthonormality}) never vanishes for some function $f(t)$ which does not produce an orthonormal basis, this gets fixed 
with ${\hat\varphi(\xi)={\hat f(\xi)}/[\sum_{n\in Z} |{\hat f}(\xi-n)}|^2]^{1/2}$. 
Once the set $\{\varphi_{0n}\}_{n\in Z}$ is orthonormal so will be the sets $\{\varphi_{jn}\}_{n\in Z}$ for all fixed $j$. 
Here, 
\begin{equation}
\varphi_{jn}(t)=2^{j/2}\varphi(2^j t -n)
\end{equation}
Each $V_j$ subspace is defined as the closure of the span of the $j$-th orthonormal set. $V_\infty$ is all of $L_2$ and $V_{-\infty}$ contains only the 0 function. 

To get a usable orthonormal basis for 
$L_2$ we need the $W_j$'s into  which the $V_{j'}$'s for $j'>j$ decompose orthogonally. We search for ``refinable'' $\varphi$'s, defined by imposing the requirement eq.  (\ref{scalinginclusion}):
\begin{equation}
\varphi_{00}(t)\equiv \varphi(t)=2^{1/2}\sum_{n\in Z} c_n \varphi(2t-n)\equiv \sum_{n\in Z} c_n \varphi_{1n}(t) \label{scalingeq}
\end{equation}

Eq. (\ref{scalingeq}) gets $\varphi(t)$ from $\varphi(2t)$ by a low pass filter acting on the index $n$. An orthogonal high pass filter will then produce the wavelet $\psi$. In signal processing one sets $c_n=0$ for $n<0$ and $n>N>0$.  $H(z)=\sum_n c_n z^{-n}$ is a polynomial in $z^{-1}$, constructed from the coefficients $h_n=2^{-1/2} c_n$. $z$ is the place holder for a shift of a function by 1 to the right.

\subsection{Compact wavelets in $D=1$}

Compactly supported wavelets obey $c_n\ne 0$ for $n=0,1,2\cdots N$ where $N>0$ is odd. The finite set of coefficients $\{c_n\}_{n=0}^N$ is fixed by certain conditions. It corresponds to a finite impulse response filter (FIR). Everything else is determined by these coefficients. To get a basis of all of $L_2$ one needs an exactly invertible filter bank transform. One finds solutions $\varphi(t)$ to eq. (\ref{scalingeq}) that have compact support with $\varphi(t)=0$ for $t<0$ and $t>N$. For each optimal set $c_n$ there is a distinct set $c_{N-n}$ when $N>1$. For all $N>1$ $\varphi(t)$ is not symmetric under $t\to N-t$ and $\varphi(N-t)$ is the solution given by the flipped coefficients $c_{N-n}$.

Eq. (\ref{scalingeq}) can be solved by iteration in Fourier space. 
\begin{equation}
\varphi(t)=2\sum_{n=0}^{n+N} h_n \varphi(2t-n)\;\Rightarrow\; \hat\varphi(\xi/2)=H(\xi/2) \hat\varphi(\xi/2),~~~H(\eta)\equiv H(e^{2\pi i \eta})
\end{equation}
The iteration produces an infinite product which has to converge. 
\begin{equation}
\hat\varphi(\xi)=\prod_{j=1}^\infty H(\xi/2^j)
\end{equation}
$H(\eta)$ has to approach 1 sufficiently rapidly when $\eta\to 0$ ($j\to\infty$). $\varphi(t)$ and $\psi(t)$ can be made differentiable only to some finite order, increasing with $N$. Differentiation beyond this order produces a rough, fractal like function. These wavelets are continuous both in time and frequency except for  the simplest prototype, the Haar basis, which is discontinuous in time. It is defined by $N=1$ and $c_0=c_1=1/2$. The Haar scaling function is
1 for $0<t<1$ and zero otherwise. Some lattice field theorists may secretly think in terms of higher dimension versions of Haar bases, viewing lattice fields as approximations of smooth functions by piecewise constant ones. The lattice continuum limit corresponds to a dynamic rather than a kinematic infinite resolution limit. 

The Haar basis is not directly useful because it is too delocalized in Fourier space. It is conceptually simpler because it is easily seen to be orthonormal. One can start an iteration of the scaling relation from the Haar scaling function but with non-Haar $c$'s and the iteration converges to the new scaling function. One finds conditions on the new $c$'s that ensure that orthonormality is inherited through the iteration process. 

To ensure orthonormality directly at fixed $j$ a condition on $H$ making eq. (\ref{orthonormality}) valid is needed. To this end a scaling relation for the relevant quantity 
$\sigma_\varphi(\xi)$ appearing there is employed. Using
\begin{equation}
\hat\varphi(2\xi+n)=H(\xi+1/2)\hat\varphi(\xi+1/2),
\end{equation}
and splitting the odd from the even contributions to the sum in the definition of $\sigma_\varphi(\xi)$ we obtain
\begin{equation}
\sigma_\varphi (2\xi) = |H(\xi)|^2 \sigma_\varphi(\xi)+|H(\xi+1/2)|^2\sigma_\varphi(\xi+1/2)
\end{equation}
This condition on $H$, equivalent to eq. (\ref{orthonormality}), is:
\begin{equation}
|H(\xi)|^2+|H(\xi+1/2)|^2=1
\end{equation}
In the time domain this is ``double-shift'' $l_2$ orthonormality of the set of infinite dimensional vectors $\{h(n-2k)\}_{k\in Z}$ at any fixed $n\in Z$. 

The wavelet $\psi$ is defined in terms of the scaling function $\varphi$ by
\begin{equation}
\psi(t)=2^{1/2}\sum_n d_n \varphi(2t-n);~~d_n=(-1)^n c_{N-n}
\end{equation}
The infinite vectors $d$ are double-shift orthonormal to each other and double-shift orthogonal to the infinite vector $d$. This ensures the full orthonormality of the set $\{\psi_{jn}\}_{j\in Z, n\in Z}$.

The upshot is that in order to produce a useful $Z^2$ lattice of basis functions of $L_2(R)$ good sets of coefficients $c_n$ need to be found. As $N$ increases $\varphi(t)$ can be made differentiable to higher orders. 

\subsection{Wavelets for $D>1$}

One generalization of wavelets to $D>1$ employs a basis made out of products of $D$ $(D=1)$-functions with  $\mu=1,2..,D$. For each $\mu$ one can pick either a $\varphi_{j,k_\mu} (x_\mu)$  
or a $\psi_{j,k_\mu}(x_\mu)$, except that the case of all factors 
being $\varphi$'s is not allowed. For each overall resolution 
$j\in Z$ and each vector ${\vec k}$ consisting of $D$-integers, one has $2^D-1$ basis elements. Let $\alpha=1,2,...(2^D-1)$ label the different elements. Notationally, one can represent $\alpha$ as a $D$-bit number $\alpha >0$. The set $\{\psi^\alpha_{j,{\vec k}}\}_{1\le\alpha\le {2^D-1}\;j\in Z\;{\vec k}\in Z^D}$ is a basis of $L_2(R^d)$. The orthonormality of the $\psi^\alpha_{j,{\vec n}}({\vec x})$ follows from 
\begin{equation}
V_{j+1}^{(D)} = \otimes_{\mu=1}^{\mu=D} ( V_{j+1} ^\mu )=\otimes_{\mu=1}^{\mu=D} (V_j^\mu \oplus W_j^\mu ) = V_{j}^{(D)}\oplus ( \oplus_{\alpha=1}^{\alpha=2^D-1}\; W_j^\alpha )
\end{equation}
where $V_j^{(D)}$ is the resolution $j$ space in $D$-dimensions and $V_j^\mu$ is the one dimensional resolution $j$ space in direction $\mu$. 

We are now in the position to add some details. For example, in $D > 2$, consider the Euclidean action 
\begin{equation}
S[\Phi]=\int d^D x \frac12 (\partial_\mu\Phi)^2 +\lambda \int d^D x (\Phi^2)^{\frac{D}{D-2}}\label{contaction}
\end{equation}
It is invariant under 
\begin{equation}
\Phi({\vec x}) \rightarrow 2^{\frac{D}{D-2}}\Phi(2{\vec x})\label{fieldscaling}
\end{equation}
The requirement $D> 2$ is there because usage of wavelets needs polynomial actions.  The kinetic energy terms is taken to be of standard form to get a familiar Gaussian fixed point at zero interaction. Standard interactions are then strictly local and have limited power growth for large fields.

Equation (\ref{fieldscaling}) corresponds in coefficient space to
\begin{equation}
\Phi^\alpha_{j{\vec n}} \;\rightarrow \; \frac{1}{2} \Phi^\alpha_{j-1\;{\vec n}}
\end{equation}
where $\Phi ({\vec x})=\sum_{j,{\vec n},\alpha}\Phi^\alpha_{j{\vec n}}\psi^\alpha_{j{\vec n}}({\vec x})$. A generic invariant multinomial of degree $k+1\ge 2 $ has the form
\begin{equation}
\sum_{j\in Z ,{\vec n}\in Z^D }2^{(k+1)j}\;\Phi^{\alpha_1}_{j{\vec n}}\; \Phi^{\alpha_2}_{j+J_1\; {\vec n}+{\vec N_1}}\cdots\Phi^{\alpha_{k+1}}_{j+J_k \;{\vec n}+{\vec N_k}}\label{genericterm}
\end{equation}
Linear terms in the action density vanish. It is also possible to include terms in the action containing derivatives of the continuum fields. One could approximately replace continuum derivative terms by finite difference terms in ${\vec n}$. Truncating the sum over $j$ in eq. (\ref{genericterm}) spoils scale invariance at the ends of the truncation interval. Translational invariance of the continuum action ${\vec x}\to {\vec x}+{\vec\Delta}$ is implemented by $j$-dependent shifts ${\vec n}\to 2^{-j}{\vec \delta}$. This is better defined for truncated $j$ ranges, where ${\vec \Delta}$ is restricted to discrete dyadic sets.

An exact rewriting of a continuum action like in eq. (\ref{contaction}) in terms of wavelet coefficients  would be complicated and dependent on the precise type of wavelets used. It seems possible, if not plausible, that simpler functionals of the $\Phi^{\alpha}_{j{\vec n}}$ would produce identical fixed points up to meaningless analytic changes of variables. These actions, when translated back to functionals of $\Phi ({\vec x})$, might not be strictly local. This may not matter. There always is the freedom to redefine fields. A continuum action will typically be invariant under $x_\mu \to -x_\mu$ for all $\mu$ independently. From it we can reconstruct a continuum action using $c_n$ or $c_{N-n}$.

There is much more to the story. Practically, explicit calculations can be carried out by iterations needing only the coefficients $c_n$. The algorithms of decomposition (analysis) of a function or the opposite direction (synthesis) are optimally efficient, beating even the FFT.

\section{What is good about the wavelet decomposition}

Wilson needed a phase-space cell decomposition because it provided a good factorized starting point for carrying out the path integral. The integrand is expressed as an infinite product over $(j,{\vec n})$ of functions of $\Phi^\alpha_{j{\vec n}}$ with only self-couplings. The inclusion of the rest of the action is under much better control than standard perturbation theory or a lattice hopping expansion. The heart of the reason is that a term like $\int [\Phi(x)]^k d^D x$ is of order $\sum_{(j,\alpha{\vec n})} [\Phi^\alpha_{(j,{\vec n})}]^k$ even for $k>2$. Scale invariance forces explicit factors of $2^j$. For quadratic terms corresponding to normal mass terms in continuum, one gets an exponential hierarchy reminiscent of deconstructed version of RS-scenarios, so maybe one can evade fine-tuning. 

\section{Supplementary remarks}

Below are some remarks that could not be fit into the time frame of the oral presentation.

\begin{itemize}
	\item Masslessness of quarks and of scalar fields are different in continuum field theory in the context of fine tuning. On the lattice they used to be closer: one needs to finely tune the fermion hopping parameter for Wilson fermions. The resolution offered by the overlap was based on exactly integrating out an infinite number of fermions, in the presence of an arbitrary lattice gauge field, carrying different flavors \cite{neub92}. This established exact chiral symmetry by using a special mass matrix in this ancillary flavor space. In \cite{gw} masslessness is ensured in the absence of gauge fields, for the simpler case of free fermions, by integrating out an infinite number of fields going from the highest momentum slices downwards. The result in, eq. (26,28) there is attributed to M. Peskin. The structure of the answer bears similarities to wavelets. This is not the right place to elaborate further. The main point is that this time it is scale invariance that is ensured by the infinite iteration. Unlike the method applied to chirality, such a method could also ensure masslessness for scalars. Using wavelets, this can be made concrete by keeping the range of $j$ doubly infinite and constructing a transfer matrix in the $j$ direction \cite{resolvingpower}. This time there is explicit $j$-dependence. It remains to be seen how this situation can be brought under control.
	
    \item In the $\epsilon$-expansion, in perturbation theory, one never needs to deal explicitly with fine-tuning the bare mass to set the physical one to zero. The reason is that action of scale transformations is neatly separated out in the formalism. The hope is that wavelets would provide a non-perturbative and more concrete 
    realization of the same effect.
	
	\item Starting from the action in terms of wavelet coefficients only, how can the wavelets that were used be recovered? One would start by identifying the dilatation and translation infinite matrices in the wavelet basis. This needs to be better understood. 
	
	\item Clearly, there is an exact lattice-$AdS$ equivalence to continuum field theory before one attempts to do the path integral. There is no gravity. At this point, one is free to dream that these two distinction from the usual $AdS/CFT$ correspondence eliminate each other in some cases.
	
	\item It is likely that for lattice field theory purposes variants of wavelet constructions, specifically tailored to the problem at hand, would be useful or even necessary.
	
	\item Wavelets are a better place to formulate a variant of lattice radial quantization \cite{radq} because translational invariance is preserved.
	
	\item For gauge theories the truncation of the range of $j$ restricts the set of gauge transformations in an awkward way. A way around this is to adopt a lattice-continuum formulation as follows: In \cite{bounds}, generalizing an observation of P. van Baal, I discussed the unitary, nonlocal, continuum operators 
	\begin{equation}
	T_{\mu\;c}=e^{aD_\mu},~~D_\mu=\partial_\mu + i A_\mu,~~\mu=1,...,D
	\end{equation}
	$a$ is a lattice spacing scale and $A_\mu$ a Hermitian matrix in the Lie Algebra of the gauge group.
	The action of $T_{\mu\;c}$ can be restricted to lattice matter fields. Here we take it as acting on  continuum matter fields. Each $T_{\mu\;c}$ can then be represented as an infinite unitary matrix acting on wavelet indices and group indices. The pure gauge action is 
	\begin{equation}\label{action}
	\sum_{\mu,\nu}^{D}{\rm Tr} \{ T_{\mu\;c},T_{\nu\;c}] [T_{\mu\;c},T_{\nu\;c}]^\dagger\} 
	\end{equation}
	The $T_{\mu\;c}$ are restricted by requiring the pure gauge action to be finite. Note that these 
	$T_{\mu\;c}$ can have non-zero entries connecting different bandwidths segments. While we are at it, we may, for gauge group $SU(N)$, take $N\to\infty$ also, putting the right power of $N$ in (\ref{action}).
	I just started exploring this formal idea.
	\item Finally, wavelets provide a new formal viewpoint related to a set of ideas circulating in the literature for some time, centered on extra dimensions and their deconstruction (discretization). In terms of wavelets we start from a continuum Lagrangian and directly transform it into a fully deconstructed (lattice) theory in one dimension higher.  
\end{itemize}

\section{Summary}

I think that wavelets should be re-examined for applications in lattice field theory. My hopes are  that they may present an alternative starting point for constructing continuum effective field theories that are free of fine-tuning problems and that useful continuum limits of the wavelet $D+1$ dimensional index space might emerge in certain situations. Also, new large $N$ master-fields may appear in this basis.


\begin{thebibliography}{99}
\bibitem{resolvingpower} H. Neuberger, ``Resolving-Power Quantization'', arXiv:1612.00023.
\bibitem{waveletsbooks} I. Debauchies, ``Ten Lectures on Wavelets'', SIAM, ISBN-13: 978-0898712742; A. Cohen, ``Numerical Analysis of Wavelet Methods'', North Holland, ISBN 0-444-51124-5; S. Mallat, ``A Wavelet Tour of Signal Processing - The Sparse Way'', Academic Press, ISBN 13: 978-0-12-374370-1; M. Jansen and P. Oonincx, ``Second Generation Wavelets and Applications'', Springer, ISBN 1-85233-916-0;
I. Ya. Novikov, V. Yu. Protasov and M. A. Skopina, ``Wavelet Theory'', Translations of Mathematical Monographs, ISBN 978-0-8218-4084-2; G. Strang and T. Nguyen, ``Wavelets and Filter Banks'', Wellesley-Cambridge Press, ISBN 0-9614088-7-1; G. Battle, ``Wavelets and Renormalization'', World Scientific, ISBN-13: 978-9810226244.

\bibitem{waveletsqft} I. G. Haliday and P. Suranyi, Nucl. Phys. B (1995) 414; C. Best, A. Sch{\" a}fer and W. Greiner, Nucl. Phys. B (Proc. Suppl.) 34 (1994) 780; C. Best, Nucl. Phys. B (Proc. Suppl.) 83-84 (2000) 848.

\bibitem{neub92} R. Narayanan, H. Neuberger, Phys. Lett. B302 (1993) 62.

\bibitem{gw} P. H. Ginsparg, K. G. Wilson, Phys. Rev. D25 (1981) 2649.

\bibitem{radq} R. C. Brower, G. T. Fleming, H. Neuberger, Phys. Lett. B721 (2013) 299; H. Neuberger, Phys. Rev. D 90  (2014) 114501.

\bibitem{bounds} H. Neuberger, Phys. Rev. D61 (2000) 085015.
\end{thebibliography}

\end{document}